\documentclass[aps,pre,reprint,10pt,superscriptaddress,nobibnotes,showpacs]{revtex4-1}
\usepackage{amsfonts}
\usepackage{amsmath}
\usepackage{amssymb}
\usepackage{graphicx}%
\usepackage{color}
\begin{document}
%\usefont{T1}{tnr}{m}{sl}

\title{ Ion-acoustic solitary waves and shocks  in a  collisional dusty negative ion plasma }
\author{A. P. Misra}
\email{apmisra@visva-bharati.ac.in; apmisra@gmail.com}
\affiliation{Department of Mathematics, Siksha Bhavana, Visva-Bharati University, Santiniketan-731 235, West Bengal, India}
\author{N. C. Adhikary}
\email{nirab$_$iasst@yahoo.co.in}
\affiliation{Physical Sciences Division, Institute of Advanced Study in Science and Technology, Vigyan Path, Paschim Boragaon, Garchuk, Guwahati-781035, Assam, India}
\author{P. K. Shukla}
\email{ps@tp4.rub.de; profshukla@yahoo.de}
\affiliation{International Centre for Advanced Studies in Physical Sciences \& Institute for Theoretical Physics,
Faculty of Physics and Astronomy, Ruhr University Bochum, D-447 80 Bochum, Germany}
\affiliation{Department of Mechanical and Aerospace Engineering \& Center for Energy Research,
 University of California San Diego, La Jolla, CA 92093, U. S. A.}
%\keywords{}
\begin{abstract}
We study the effects of ion-dust collisions and   ion kinematic viscosities on the linear ion-acoustic instability as well as the nonlinear propagation of small  amplitude solitary waves and shocks (SWS) in a   negative ion plasma with immobile charged dusts.      {The  existence  of two linear ion  modes, namely the `fast' and `slow' waves is shown, and their properties are analyzed in the  collisional negative ion plasma.}    {Using the standard reductive perturbation technique,  we derive a modified Korteweg-de Vries-Burger (KdVB) equation which describes the evolution of  small  amplitude SWS.}   {The    profiles of the latter are numerically examined   with parameters relevant for laboratory and space plasmas where charged dusts may be   positively or negatively charged.} It is found that negative ion plasmas containing positively charged dusts support the  propagation of  SWS with negative potential. However, the perturbations with both positive and negative potentials may exist when dusts are negatively charged.  The results may be useful for the excitation of  SWS in laboratory negative ion  plasmas as well as for observation in space plasmas where charged dusts may be positively or negatively charged.
\end{abstract}
%\received{05 September 2012}
 %\revised{14 October 2012}
\maketitle
\section{Introduction} 
 The nonlinear propagation of solitary waves and shocks (SWS)   in dusty plasmas have been widely studied   for understanding the electrostatic disturbances in space plasma environments \cite{1,2} as well as in laboratory plasma devices \cite{3,4}.  Many researchers have   pointed out that  charged dust grains can drastically modify the existing response of electrostatic wave spectra in plasmas depending upon whether the charged dusts are considered to be static or mobile \cite{3,4,5,6,7,8,9,10,11,12}. 
 
The existence of dust ion-acoustic (DIA) solitons was  first predicted by  Shukla and Silin \cite{6}, and  was later observed experimentally by Barkan \textit{et al} \cite{3}. The phase velocity of these DIA waves increases when the density of electrons decreases. The nonlinear evolution of such small amplitude   waves   is described by  the  Korteweg de-Vries (KdV) equation, which was first derived by Washimi and Tanuiti \cite{14} in a   two-component plasma. It was also reported that the KdV equation can have both compressive and rarefactive solitary wave solutions  when   sufficient amount of negative ions are present in multi-component plasmas \cite{15}. Furthermore, it was also observed that when the negative ion density exceeds a critical value, only negative or rarefactive solitons can be excited \cite{experiment-negative-ion}. 

One of the nonlinear phenomena of DIA waves is the generation of  shocks,   which have already been observed experimentally \cite{16}.  The shock waves are described with an additional Burger term due to dissipation in the KdV equation, known as the KdV-Burger (KdVB) equation \cite{4,16}. The nonlinear properties of   dust ion-acoustic shocks and holes have been studied \cite{17} with the  consideration of ion kinematic viscosity in a dusty plasma. Furthermore, the nonlinear properties of small amplitude ion-acoustic shocks in a collisional dusty plasma has been investigated by  Ghosh \textit{ et al} \cite{18}. In their analysis, they have ignored the frictional force due to ion-dust collision and considered the effects of ion kinematic viscosity in dusty plasmas with positive ions.  Mamun \textit{et al} \cite{11} investigated the existence of dust electron-acoustic shocks in  a  negative ion plasma with dust charge fluctuation.  They  reported the formation of DIA shocks with negative potentials in the plasma.  In a recent work of Adhikary \cite{12}, the effect of kinematic viscosity has been considered to study the propagation of DIA shocks in multi-ion plasmas, and it has been stressed  that the viscosity in dusty plasmas plays   a key role in the formation of DIA shocks. Thus, both the viscosity and collision of ions with dusts   may play crucial roles in dissipation on the propagation of SWS in dusty multi-ion plasmas.
 
Since the charging rates of dusts by the positive and negative ions are nearly balanced, the existence and the formation of dusty plasmas with both positive and negative ions may be possible   under some circumstances.   Recently, Kim \textit{et al} \cite{19-Kim-Merlino}  have investigated that dust particles injected  in laboratory negative ion plasmas can become positively charged  when the number density of negative ions greatly exceeds ($\gtrsim500$) that of the electrons. In space environments, the possible role of negative ions has been discussed, and it has been found that  dusts can be positively charged if there is a sufficient number density of heavy negative ions (with mass $\gtrsim300$ amu) \cite{20-Rapp}.
 
In this  work, we investigate the ion-acoustic instability as well as the  nonlinear propagation   of   DIA SWS in an unmagnetized collisional dusty plasma containing both   positive and negative ions. The effects of ion thermal pressures, ion-dust collisions as well as the kinematic viscosity of the ion fluids are taken into account to maintain the equilibrium of ions. The present paper generalizes and extends the previous work of Adhikary \cite{12} to include collisional effects of ions with dusts, different mass    and thermal pressures of ions as well as different expression for ion kinematic viscosities. We show that when the negative ion concentration is much larger than the density of electrons or when plasma is the admixture of positively charged (static) dusts, SWS with negative potential exist. However, when the concentration of negative ions is smaller than a certain value or plasma contains negatively charged dusts, both the compressive and rarefactive SWS may exist.
\section{Basic equations}
We consider the one-dimensional propagation of DIA waves in an unmagnetized collisional dusty plasma, which consists of singly charged  adiabatic positive and negative ions, Boltzmann distributed electrons and immobile charged dusts. Since the dusts are too heavy to move on the time scale of the ion-acoustic waves, we do not consider the dynamics of charged dusts. The latter can, however, affect the collision rates with  ions as well as the wave dispersion and nonlinearity. We have     neglected the electron inertia, since the electron thermal speed is much larger than that of ions. We also assume that  the negative ions are heavier than the positive ions.    The immobile dust particles carry some charges so as to maintain the overall charge neutrality condition given by
\begin{equation}
n_{e0}+n_{n0}=n_{p0}\pm z_dn_{d0}, \label{charge-neutrality-1}
\end{equation} 
where $n_{j0}$ is the unperturbed number density of charged species $j$ ($j=e$, $p$, $n$, $d$, respectively, stand for electrons, positive ions, negative ions and static dusts), $z_d$ $(>0)$ is the dust charge state. The upper (lower) sign in Eq. \eqref{charge-neutrality-1} corresponds to positively (negatively) charged   dusts.  The condition  \eqref{charge-neutrality-1}   can also be written as
\begin{equation}
\mu_e=\mu_i+\zeta\mu_d-1, \label{charge-neutrality-2}
\end{equation}
where $\mu_e=n_{e0}/n_{n0}$,  $\mu_i=n_{p0}/n_{n0}$, $\mu_d=z_dn_{d0}/n_{n0}$ are the density ratios and  $\zeta=1$ $(-1)$ correspond  to positively (negatively) charged dusts. 
 The basic equations in one space dimension are
\begin{equation}
\frac{\partial n_j}{\partial t}+\frac{\partial}{\partial x}(n_j v_j)=0, \label{cont-eq}
\end{equation}
\begin{equation}
\left(\frac{d}{dt}+\nu_{jd}\right)v_j=-\frac{q_j}{m_j}\frac{\partial \phi}{\partial x}-\frac{3k_BT_j}{2m_jn^2_{j0}}\frac{\partial n^2_j}{\partial x}+\eta_{j}\frac{\partial^2 v_j}{\partial x^2}, \label{moment-eq}
\end{equation}
\begin{equation}
\frac{\partial^2 \phi}{\partial x^2}=4\pi e\left(n_e-n_p+n_n-\zeta z_dn_{d0}\right),\label{poisson-eq}
\end{equation}
and the Boltzmann distribution for electrons 
\begin{equation} 
n_e=n_{e0}\exp\left({e\phi}/{k_BT_e}\right)\label{electron-eq},
\end{equation}
 where $d/dt=\partial_t+v_j\partial_x$ is the convective derivative. The physical quantities $n_j$, $v_j$ and $m_j$ respectively denote the number density, velocity and mass of $j$-species particles. Furthermore, $q_p=e$ and $q_n=-e$, where $e$ is the elementary charge. Also, $\phi$ is the electrostatic potential, $k_B$ is the Boltzmann constant and $T_j$ is the particle's thermodynamic temperature. In Eq. \eqref{moment-eq},  $\nu_{jd}$ $(\sim\sigma_{jd}n_{d0}v_{tj})$ denotes the  collision rate in which $\sigma_{jd}$ is the collision cross section  of $j$-species ions with   dusts    and $v_{tj}=\sqrt{k_BT_j/m_j}$ is the thermal velocity of $j$-species ions.  Moreover,    $\eta_{j}$ is the  kinematic viscosity of $j$-species ions which arises mainly due to the ion-dust collisions. We here mention that the exact expression for $\eta_{j}$ in terms of    $\nu_{jd}$ or $n_{d0}$ has not yet fully investigated so far. So, we will consider arbitrary values of it  in order to see the damping of DIA waves numerically \cite{Nakamura-Sarma}.    We have used the adiabatic equation of state in Eq. \eqref{moment-eq}, namely $P_j/P_{j0}=(n_j/n_{j0})^3$ with $P_{j0}=n_{j0}k_B T_j$ for each of the ion species.  The adiabatic index $\gamma=3$ $[=(2+D)/D$, $D$ being the number of degrees of freedom] is due to the one-dimensional geometry of the system. Furthermore, in the ion continuity equations  the electron-ion recombination effects, being smaller at low-pressure limit ($\sim4-5\times10^{-4}$ Torr), have been neglected \cite{16}. However, the contribution from the frictional force $\nu_{jd}v_j$ may not be neglected compared to that from the viscous term $\varpropto\eta_j$.  {It has been found that in laboratory experiments \cite{Nakamura-Sarma},  the nondimensional viscosity parameter $\tilde{\eta}_j=\eta_j/\lambda_D^2\omega_{pn}$ may be about $10$ times of the collisional frequency $\tilde{\nu}_{jd}=\nu_{jd}/\omega_{pn}$, where $\omega_{pn}=\sqrt{4\pi n_{n0}e^2/m_n}$ is the negative ion plasma frequency and $\lambda_D=\sqrt{k_B T_e/4\pi n_{n0}e^2}$ is the Debye length.  Also,  $\tilde{\eta}_j$ and  $\tilde{\nu}_{jd}$ are almost linearly proportional to the dust number density $n_{d0}$.}

Next, we normalize the physical quantities according to
$\phi\rightarrow e\phi/k_BT_e$, $n_j\rightarrow n_j/n_{j0}$, $v_j\rightarrow v_j/c_s$, where $c_s=\sqrt{k_BT_e/m_n}$ is the ion-acoustic speed. The space and time variables are normalized by the Debye length $\lambda_D$  and the inverse of the negative ion  plasma frequency $\omega_{pn}$  respectively. Thus, from Eqs. \eqref{cont-eq}-\eqref{poisson-eq} we obtain the following normalized set of equations 
\begin{equation}
\frac{\partial n_j}{\partial t}+\frac{\partial}{\partial x}(n_j v_j)=0, \label{Ncont-eq}
\end{equation}
\begin{equation}
\left(\frac{d}{dt}+\tilde{\nu}_{jd}\right)v_j=\beta_j\left(\mp\frac{\partial \phi}{\partial x}-\frac{3}{2}\sigma_j\frac{\partial n^2_j}{\partial x}\right)+\tilde{\eta}_{j}\frac{\partial^2 v_j}{\partial x^2}, \label{Nmoment-eq}
\end{equation}
\begin{equation}
\frac{\partial^2 \phi}{\partial x^2}=\mu_e e^{\phi}-\mu_in_p+n_n-\zeta\mu_{d},\label{Npoisson-eq}
\end{equation}
where  the upper (lower) sign in $\mp$ on the right-hand side of Eq. \eqref{Nmoment-eq}  corresponds to positive (negative) ions, $\sigma_j=T_j/T_e$,  $\beta_j=m_n/m_j$  with $\beta_p\equiv\beta$, say. 
\section{Dispersion relation: ion-acoustic instability}
In order to identify the different  wave eigenmodes and to study their stability/ instability under the dissipation due to ion-dust collisions and ion kinematic viscosities, we perform a linear analysis in which    the perturbations of the physical quantities vary as   $f\sim\exp(ikx-i\omega t)$, where $\omega$ $(k)$ is the wave frequency (number). Thus, Fourier analyzing the  Eqs. \eqref{Ncont-eq}-\eqref{Npoisson-eq} we obtain the following  {dispersion relation
\begin{equation}
D(\omega,k)\equiv1+\sum_{j=e,p,n}\chi_j=0,\label{Dispersion-R}
\end{equation}
where $\chi_e=\mu_e/k^2$ and}
\begin{equation}
\chi_p=\mu_i \beta\left[3\beta\sigma_pk^2-\omega\left(\omega+i\tilde{\nu}_{pd}+i\tilde{\eta}_pk^2\right)\right]^{-1}, \label{Disp-exp1}
\end{equation}
\begin{equation}
\chi_n=\left[3\sigma_nk^2-\omega\left(\omega+i\tilde{\nu}_{nd}+i\tilde{\eta}_nk^2\right)
\right]^{-1}. \label{Disp-exp2}
\end{equation}
We note that the damping terms $i(\tilde{\nu}_{jd}+\tilde{\eta}_jk^2)$ for $j=p$ and $n$ arise due to the  ion-dust collisions  and the viscosity effects, and thereby cause the DIA instability in the collisional plasma. When $k^2\ll1$, $\tilde{\eta}_jk^2$ term is negligible compared to $\tilde{\nu}_{jd}$. Thus, for long wavelength modes, the damping effect is mainly due to the ion-dust collisions.  In particular, in absence of   these dissipative effects, the dispersion relation Eq. \eqref{Dispersion-R} reduces to \cite{experiment-negative-ion}
\begin{equation}
1+\frac{\mu_e}{k^2}=\frac{\mu_i\beta}{\omega^2-3k^2\beta\sigma_p}+\frac{1}{\omega^2-3k^2\sigma_n}.\label{Disp1}
\end{equation}
The first and second terms on the right-hand side of Eq. \eqref{Disp1} are,  respectively, the contributions from the positive and negative ion species, whereas the term $\varpropto\mu_e$ is from the electron species in plasmas and the constant term $1$ is from the effect of dispersion due to charge separation (deviation from quasineutrality) of the species.  If the phase speed of the wave satisfies $\omega/k\gg c_s>v_{tj}$, then   from Eq. \eqref{Disp1} the fast ion-acoustic wave mode can be recovered as
\begin{equation}
\omega^2\approx\frac{(1+\mu_i\beta)k^2}{\mu_e+k^2}. \label{Disp_1}
\end{equation}
In absence of   negative ions in plasmas one obtains from Eq. \eqref{Disp1}   the following disperson relation \cite{6}. 
\begin{equation}
\omega^2= 3\beta\sigma_p k^2+\frac{\mu_i\beta k^2}{\mu_e+k^2}. \label{Disp2}
\end{equation}
From Eqs. \eqref{Disp_1} and \eqref{Disp2} we find that for a fixed $\beta$, $\mu_i$ and/or $\sigma_p$,  the frequency of the ion wave modes increases with the wave number. However, as the wave number increases beyond a certain value, e.g., $k\gtrsim0.2$, the fast mode in the limit may approach more or less a constant  value.  Also, as the dust number density increases, i.e., $\mu_e$ increases (decreases) for positively (negatively) charged dusts, the phase velocity of the fast mode decreases (increases). Physically, this decrease (increase) of the phase velocity is attributed to the addition (removal) of electrons by the dust grains in order to maintain the charge neutrality condition \eqref{charge-neutrality-1} of the plasma. 
\begin{figure*}
\includegraphics[height=4in,width=6in]{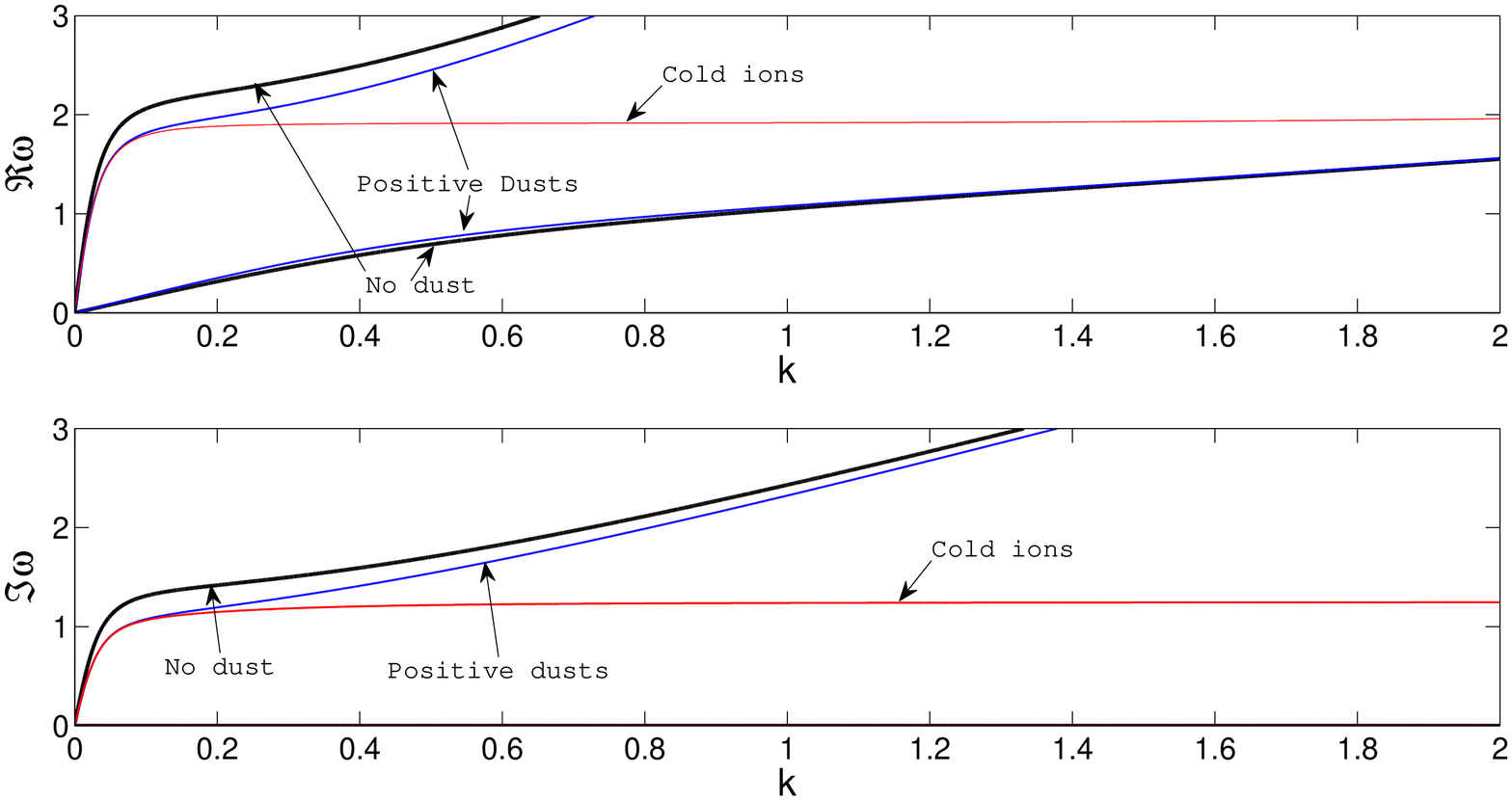}  
\caption{(Color online) The Real (upper panel) and imaginary (lower panel) parts of $\omega$ versus $k$ obtained as a numerical solution of the dispersion relation [Eq. \eqref{Dispersion-R}]. The thin (blue) and thick (black)  lines correspond to the cases when positively charged dusts are present  and when no dust is present in the plasma respectively. The line (red) which asymptotically approaches to more or less a constant value with $k$ is corresponding to the case when ions are cold. The parameters are  for laboratory negative ion plasmas with positively charged dusts and given by $m_n=146m_{\text{prot}}$, $m_p=39m_{\text{prot}}$ ($m_{\text{prot}}$ is the proton mass);  $T_e\sim T_p\sim0.2$ eV, $T_n\sim T_e/8$, $n_{n0}\sim2\times10^9$ cm$^{-3}$, $n_{e0}=n_{n0}/700$, $n_{p0}=500n_{e0}$, $n_{d0}\sim1.7\times10^6$ cm$^{-3}$, $\tilde{\eta}_p=0.2$ and $\tilde{\eta}_n=0.1$.} 
\end{figure*}

We numerically analyze the dispersion relation Eq. \eqref{Dispersion-R} by considering parameters that may be representative of laboratory negative ion plasmas with positively and negatively charged dusts. As in Ref. \cite{PSS}, we consider a plasma in which positive ions are singly ionized $K^+$ and the heavy negative ions are $SF_6$, such that $\beta\equiv m_n/m_p=146/39\approx3.74$. The temperature and density are considered as $T_e\sim T_p\sim0.2$ eV, $T_h\sim T_e/8$ with $n_{n0}\sim2\times10^9$ cm$^{-3}$, $n_{e0}=n_{n0}/700$, $n_{p0}=500n_{e0}$ for positively charged dusts  (and $n_{e0}=n_{n0}/100$, $n_{p0}=150n_{e0}$ for negatively charged dusts)   to a surface potential $\phi_s\sim0.1$ V and grain radius $R=5\mu$m, so that $z_d\sim R\phi_s/e\sim350$. In this case, $n_{d0}\sim1.7\times10^6$ cm$^{-3}$ for which the rate of ion collisions with dusts would be higher than that with neutrals. We identify  two ion wave modes, namely the `fast' and `slow' modes as exhibited in Figs. 1-3.

Figure 1 shows the plots of the real wave frequency (upper panel) and the growth rate (lower panel) of the DIA instability  versus the wave number $k$ obtained by numerically solving Eq. \eqref{Dispersion-R} for positively charged dusts. The thick (black) line corresponds to the case when no dust is present in the plasma, whereas the thin (blue) line is for the case when  dusts (positively  charged) are present. In absence of ion thermal pressures, the real wave frequency and the growth rate assume almost a constant value for $k\gtrsim0.2$ as shown in Fig. 1 (red lines). This is evident from the analytic expression \eqref{Disp_1}. It is seen that the growth  occurs, however, the frequency is reduced in presence of dusts. Thus, when dust is introduced into the plasma, the enhanced ion-dust collision frequency damps the instability rate.  From Fig. 2, we observe that  when the dusts are negatively charged, the real wave frequency and the growth rate become larger than that the case of no dust.   This is in contrast to the case of positively charged dusts. The behaviors of these modes with or without charged dusts in plasmas suggest   a possible diagnostic for the roles of positively and negatively charged dusts as well as   the effects of ion-dust collisions.   
\begin{figure*}
\includegraphics[height=4in,width=6in]{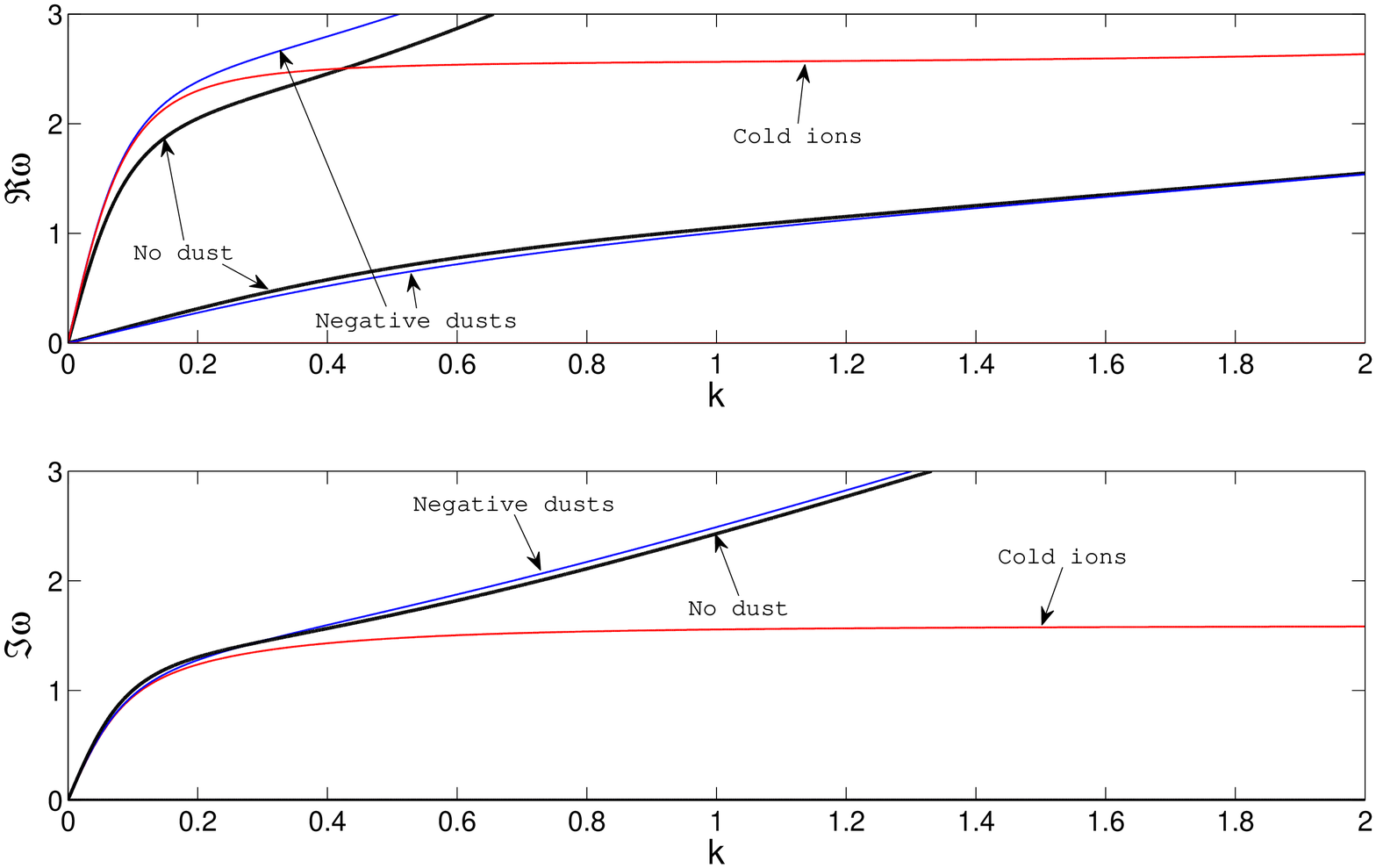}  
\caption{(Color online) The Real (upper panel) and imaginary (lower panel) parts of $\omega$ versus $k$ obtained as a numerical solution of the dispersion relation [Eq. \eqref{Dispersion-R}]. The thin (blue) and thick (black)  lines correspond to the cases when negatively charged dusts are present  and when no dust is present in the plasma respectively. The line (red) which asymptotically approaches to more or less a constant value with $k$ is corresponding to the case when ions are cold. The parameters are  for laboratory negative ion plasmas with negatively charged dusts and given by  $n_{n0}\sim2\times10^9$ cm$^{-3}$, $n_{e0}=n_{n0}/100$, $n_{p0}=150n_{e0}$ and  $n_{d0}\sim2.8\times10^6$ cm$^{-3}$. Other parameters are the same as in Fig. 1.  } 
\end{figure*}
 
Recently, it has been suggested by Rapp \textit{et al}   \cite{20-Rapp} that  the presence of sufficient amount $(n_{n0}>40n_{e0})$ of heavy negative ions (with mass $>300$ amu) may play important roles in the observation of positive dusts in the Earth's mesosphere between about $80$ to $90$ km.  They also remarked that the presence of negative ions corresponding to sub-nm size ($0.3-0.4$ nm) negatively charged dusts may lead to the positive dusts with larger size ($\sim2$ nm) grains.  In order to observe some features of the DIA modes in space plasmas, we consider plasma parameters that are representative of a dusty region at an altitude of about $95$ km where  $m_n/m_p=300/28\approx10.7$,  $T_e\sim T_p\sim T_n\sim200$ K,  $n_{n0}\sim10^6$ cm$^{-3}$, $n_{p0}\sim2\times10^4$ cm$^{-3}$ and $n_{d0}\sim10^6$ cm$^{-3}$. We assume that there is a population of sub-nm size negatively charged dusts in a dusty meteor trail (of radius $R=0.4$ nm) region in the upper atmosphere. The features of the wave modes are shown in Fig. 3 for two different values of the density ratios: $n_{n0}/n_{e0}=10$ (thin line) and  $n_{n0}/n_{e0}=50$ (thick line). We find that the wave frequency and the growth rates are reduced by increasing the ratio. The qualitative behaviors of the other mode and the effects of charged dusts and the thermal pressure remain similar as Figs. 1 and 2. In this case, we also note that  the frequency and the growth rates increase faster with $k$ having  larger magnitudes compared   to those in the case of laboratory plasmas described above. If such   instabilities can occur, VHF/UHF radar scattering from DIA waves may be possible diagnostic for the presence of sufficient amount of negative dusts in such meteor trail regions \cite{19-Kim-Merlino,PSS}. 
\begin{figure*}
\includegraphics[height=4in,width=6in]{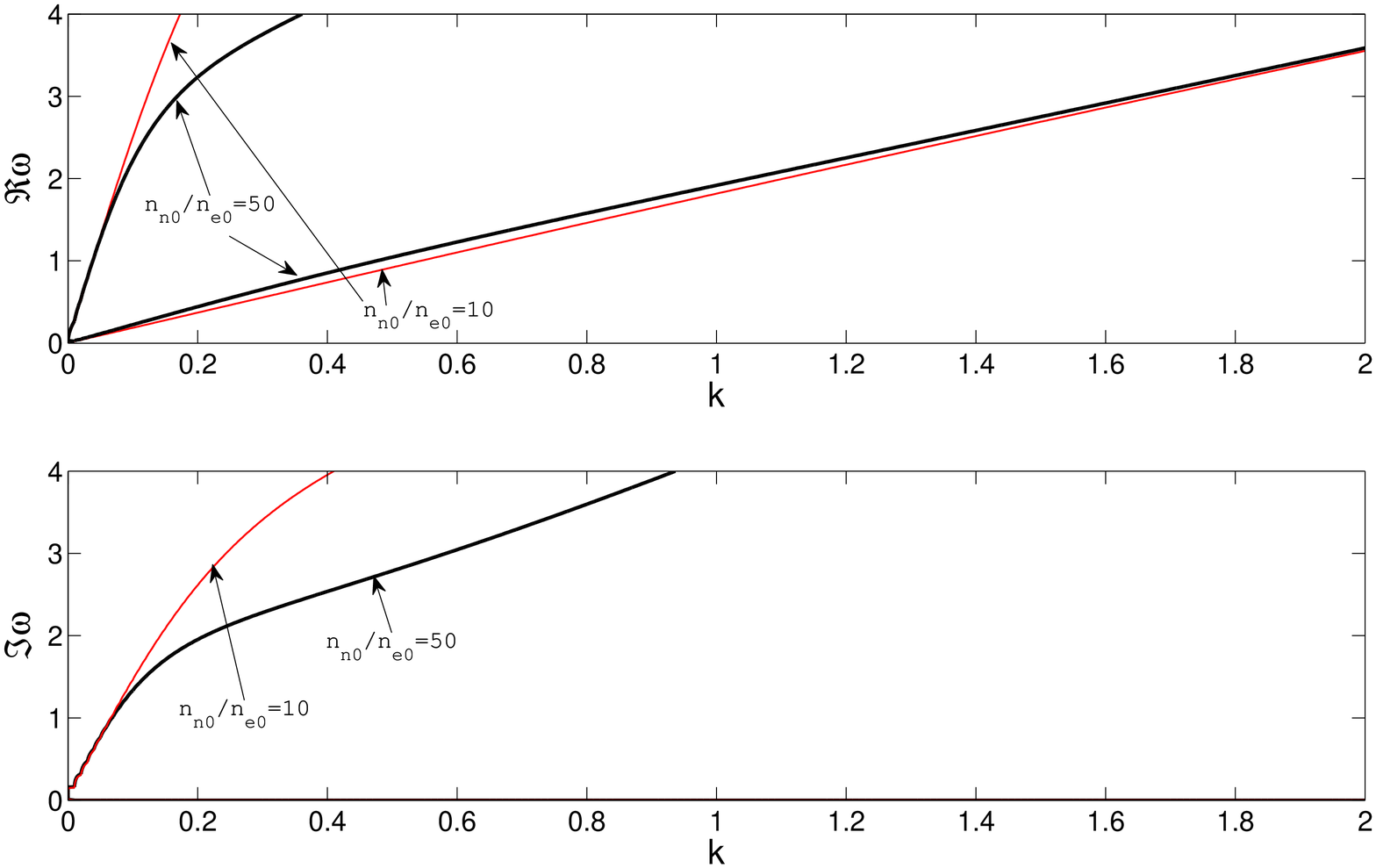}  
\caption{ (Color online) The Real (upper panel) and imaginary (lower panel) parts of $\omega$ versus $k$ obtained as a numerical solution of the dispersion relation [Eq. \eqref{Dispersion-R}]. The thin (red) and thick (black)  lines correspond to the two values of the density ratio: $n_{n0}/n_{e0}=10$ and $50$ respectively.  The parameters are  for space plasmas with negatively charged dusts and given by $m_n=300m_{\text{prot}}$, $m_p=28m_{\text{prot}}$ ($m_{\text{prot}}$ is the proton mass);  $T_e\sim T_p\sim T_n\sim 200$ K, $n_{n0}\sim10^6$ cm$^{-3}$,  $n_{p0}=60n_{e0}$ and $n_{d0}\sim10^6$ cm$^{-3}$.  } 
\end{figure*}

It has been observed in experiments that slow wave modes do not favor the formation of solitons \cite{experiment-negative-ion,experiment-Wong}.  So,   in the nonlinear regime,   the fast  mode  may propagate as DIA solitary waves due to nice balance of the dispersive and nonlinear effects. However, as the dissipation effects due to ion-dust collisions and ion kinematic viscosities, are entered into the dynamics, the system's evolution shows shock-like perturbations.  These features will be studied in the next section.
\section{Derivation of the evolution equation}
We consider the nonlinear propagation of   small but finite amplitude   DIA waves in a collisional dusty negative ion plasma. In order to observe the evolution of the waves in a different frame of reference which is moving with a speed $v_0$, we  stretch the coordinates as $\xi=\epsilon^{1/2}(x-v_0t)$ and $\tau=\epsilon^{3/2}t$,  where $\epsilon$ is a small parameter measuring the weakness of perturbations.  We also assume that $\tilde{\nu}_{jd}=\epsilon^{3/2}\nu_{j0}$ and $\tilde{\eta}_{j}=\epsilon^{1/2}\eta_{j0}$, where $\nu_{j0}$ and $\eta_{j0}$ are of the order of unity or less. Such a consideration of smallness of $\eta_{j}$ and   $\nu_{jd}$ can be found in the literature \cite{18}, and is valid in many experimental situations  (see e.g., Ref. \cite{Nakamura-Sarma}).    The dynamical variables are expanded as  
\begin{eqnarray}
&&n_j=1+\epsilon n_j^{(1)}+\epsilon^2 n_j^{(2)}+\cdots,\notag \\
&&v_j=\epsilon v_j^{(1)}+\epsilon^2 v_j^{(2)}+\cdots,\label{expansion-nvphi} \\
&&\phi=\epsilon \phi^{(1)}+\epsilon^2 \phi^{(2)}+\cdots.\notag \\
\end{eqnarray}
We then substitute these expansions   and the stretched coordinates into   Eqs. \eqref{Ncont-eq}-\eqref{Npoisson-eq}, and equate different powers of $\epsilon$. In the lowest order of $\epsilon$ (i.e., $\epsilon^{3/2}$) we obtain the following relations for the first order perturbations
\begin{equation}
n_j^{(1)}=\alpha_j \phi^{(1)},\hskip5pt v_j^{(1)}= \alpha_jv_0 \phi^{(1)}, \label{1st-pertur} 
\end{equation}
together with the expression for the wave speed in the moving frame of reference as 
\begin{equation}
v^2_0=\frac{1}{2\mu_e}\left[S\pm \sqrt{S^2-12\beta \mu_e\left[\sigma_p+\sigma_n(\mu_i+3\mu_e\sigma_p) \right]}\right], \label{phase-velo}
\end{equation}
 where   $S=1+\mu_i\beta+3(\sigma_n+\beta\sigma_p)\mu_e$ and $\alpha_j=\pm\beta_j/(v^2_0-3\beta_j\sigma_j)$. Here  $\pm$ denotes the quantities corresponding to positive $(j=p)$ and negative $(j=n)$  ions respectively. This $\pm$ sign appears due to the consideration of different mass and temperatures of positive and negative ions. However, in a limited situation  where $m_p=m_n$ and $T_p=T_n$, we can recover the  expression (here one should consider the plus sign before the square root in $v_0^2$) as in Ref. \cite{12}. However, in the next section, we will   see that only the minus sign (of $\pm$)    favors the formation of SWS in laboratory and space plasmas, since for the plus sign,  the nonlinear term becomes much larger than the dispersive term in the evolution equation to be shown shortly. 
  
Proceeding to the next order of $\epsilon$ (i.e., $\epsilon^{5/2}$) we obtain the following set of equations for the second order perturbed quantities
\begin{equation}
-v_0\frac{\partial n^{(2)}_j}{\partial\xi}+\alpha_j\frac{\partial \phi^{(1)}}{\partial\tau}+\alpha_j^2v_0\frac{\partial \left(\phi^{(1)}\right)^2}{\partial \xi}+\frac{\partial v^{(2)}_j}{\partial \xi}=0,\label{2nd-order-cont}
\end{equation}
\begin{eqnarray}
&&\left(\frac{\partial}{\partial \tau}+\tilde{\nu}_{j0}\right)\alpha_jv_0\phi^{(1)}+\frac{1}{2}\alpha_j^2\left(v_0^2+3\beta_j\sigma_j \right)\frac{\partial \left(\phi^{(1)}\right)^2}{\partial \xi}\notag \\
&&=v_0\left(\frac{\partial v^{(2)}_j}{\partial\xi}+\eta_{j0}\alpha_j\right) \frac{\partial^2 \phi^{(1)}}{\partial\xi^2} \notag \\
&&-\beta_j\left(3 \sigma_j \frac{\partial n_j^{(2)}}{\partial\xi}\pm \frac{\partial \phi^{(2)}}{\partial\xi}  \right), \label{2nd-order-moment}
\end{eqnarray}
\begin{eqnarray}
&&\frac{\partial^3 \phi^{(1)}}{\partial\xi^3}=\mu_e\left(\frac{1}{2}\frac{\partial \left(\phi^{(1)}\right)^2}{\partial \xi}+\frac{\partial \phi^{(2)}}{\partial\xi}\right)+\frac{\partial n_n^{(2)}}{\partial\xi}\notag \\
&&-\mu_i\frac{\partial n_p^{(2)}}{\partial\xi}, \label{2nd-order-poisson}
\end{eqnarray}
where  $\pm$ sign in Eq. \eqref{2nd-order-moment}   corresponds  to positive $(j=p)$ and negative $(j=n)$  ions respectively. 

Eliminating the second order quantities from Eqs. \eqref{2nd-order-cont}-\eqref{2nd-order-poisson}, we obtain, after few steps, the following evolution equation of the Korteweg de-Vries-Burger (KdVB) type.
\begin{equation}
\frac{\partial\Phi}{\partial\tau}+A\Phi\frac{\partial\Phi}{\partial\xi}+B\frac{\partial^3\Phi}{\partial\xi^3}=\eta \frac{\partial^2\Phi}{\partial\xi^2}-\nu \Phi,\label{KdVB-eq}
\end{equation}
where $\Phi\equiv\phi^{(1)}$. The coefficients of  nonlinearity and dispersion as well as  of dissipation  due to the ion kinematic viscosities and the ion-dust collisions   are,  respectively, given by
\begin{equation}
A=\frac{3\alpha_p^3\mu_i(v^2_0+\beta\sigma_p)+3\beta\alpha_n^3(v_0^2+\sigma_n)+
\beta(\alpha_n-\mu_i\alpha_p)}{2v_0(\mu_i\alpha_p^2+\beta\alpha_n^2)},\label{nonlinear-coeff}
 \end{equation}
\begin{equation}
B=\frac{\beta}{2v_0(\mu_i\alpha_p^2+\beta\alpha_n^2)}, \label{dispersion-coeff}
\end{equation}
\begin{equation}
\eta=\frac{\mu_i\eta_{p0}\alpha_p^2+\beta\eta_{n0}\alpha_n^2}{2(\mu_i\alpha_p^2+\beta\alpha_n^2)}, \hskip5pt \nu=\frac{\mu_i\nu_{p0}\alpha_p^2+\beta\nu_{n0}\alpha_n^2}{2(\mu_i\alpha_p^2+\beta\alpha_n^2)}.\label{dissip-colli-coeff}\\
\end{equation}
In particular, for $m_n=m_p$, $T_n=T_p$ and $\eta_n=\eta_p$, we recover the expressions as in Ref. \cite{12}.  When the dissipation terms proportional to $\eta$ and $\nu$ are ignored, i.e., in absence of frictional force and viscous stress, the resultant equation is the  KdV equation discussed in Ref. \cite{experiment-negative-ion}. Thus, the Burger term in Eq. \eqref{KdVB-eq} gives rise the generation of shocks and the frictional force term provides damping of the wave due to ion-dust collision. The latter increases with the increase of dust number density. 

\section{Results and Discussion}
Typically, the existence of  SWS with positive or negative potential   depends on the sign of the nonlinear coefficient $A$. So, we  numerically examine the behaviors of the  nonlinear and the dispersion coefficients $A$ and $B$ with the variations of the positive to negative ion density ratio $\mu_i$. The parameters are considered for both positively (Fig. 4) and negatively (Fig. 5) charged dusts as in Fig. 1. The solid and dashed lines correspond to $T_n=T_e/8$, $T_p\sim T_e$  and $T_n=T_e/10$, $T_p\sim T_e/2$   respectively. We have considered, respectively,    the negative and positive signs in the expression for  $v_0^2$  [Eq. \eqref{phase-velo}] for  the subplots [(a), (c)] and [(b), (d)] of Figs. 4 and 5 to compare    the corresponding values of $A$ and $B$.  {From Figs. 4(b) and (d) we find that for  multi-ion plasmas with positively charged dusts,  the numerical values of the dispersion coefficient $B$ are much larger than those of the nonlinear coefficient $A$. However, for plasmas with negatively charged dusts, the coefficient $A$ can become larger than $B$ with an increasing value of $\mu_i$ [See subplots 5(a) and (c)].  So, in both these cases, the balance between the nonlinear and the dispersive effects   may not occur for the formation of   KdV solitons  in collisional dusty negative ion plasmas.  Furthermore, when the dissipation overwhelms the wave dispersion and it is  in nice balance with the nonlinearity arising from the nonlinear mode coupling of finite amplitude waves, the monotonic shocks may be generated.  }  It turns out that  the values of $A$ and $B$ in the subplots [(a), (c)] of Fig. 4  may favor the formation of SWS with negative potentials  {(Since $A$ maintains only the negative sign)}, whereas those   in the subplots [(b), (d)] of Fig. 5 may favor the formation of SWS with both positive and negative potentials  {(Since $A$ can have both the positive and negative signs)}.  Thus, we may conclude that  plasmas with positively charged dusts may support the propagation of SWS with negative potentials, however, SWS with both positive and negative potential may exist when dusts are negatively charged.  In the former case, the number density of negative ions   greatly exceeds that of electrons and positive ions, whereas, positive ion density is to be larger than the density of  negative ions in  the latter case.  In our numerical solution of Eq. \eqref{KdVB-eq} below, we present  the characteristic features of those SWS  which have   negative potential (as the qualitative features of SWS with positive potential will be similar). 

Next,  we numerically solve Eq. \eqref{KdVB-eq} by Runge-Kutta scheme with an initial condition of the form $\Phi(\xi)=-0.03\text{sech}^2(\xi/15)$.  The development of this waveform at different times and for different parameters are shown in Figs. 6-8. The parameter values are considered as in Figs. 1-3    for both positively and negatively charged dusts in electronegative plasmas.  Figure 6 shows the case of KdV soliton, i.e., in absence of any dissipation for different times: (a) $\tau=0$, (b) $\tau=200$, (c) $\tau=300$ and (d) $\tau=500$. We see that  the leading part of the initial pulse steepens  due to positive nonlinearity. As the time progresses, the pulse separates into solitons and a residue due to the wave dispersion. It is clear from Fig. 6(d) that once the solitons are formed and separated, they propagate without changing their shape  due to the nice balance of the nonlinear and dispersion terms.  This subplot also shows that (see the dashed line) when charged dusts are introduced into the plasma,  the magnitude of the soliton height is reduced, but the   width is increased. However, as the number density of negative ion increases (see the dotted line), both the amplitude and width of the soliton decrease. In each of these cases the soliton is seen to be up-shifted.
 
Figure 7 shows the profiles of the solution of KdVB equation for plasmas with positively charged dusts. In this case,  the values of $\eta_{j0}$ and $\nu_{j0}$ are considered arbitrarily in order to observe the damping of solitary waves \cite{Nakamura-Sarma}. From Fig. 7(b), it is clear that as the dissipation due to ion kinematic viscosity enters into the dynamics, the solitary pattern breaks up and a shock profile is formed. Also, as    the   term  $\varpropto\eta_j$ increases,  the height of the shock profile decreases. Physically, the increase of $\eta_j$ corresponds to the increase of the dust number density, which, in turn, increases the ion-dust collision frequency. While compared with Fig. 7 (a), Figures 7(c) and (d) show the wave damping due to the presence of ion-dust collisions. The latter are shown to decrease the wave amplitude significantly as in Fig. 7(d) when some larger values of them are considered. From Figure 7(c) we find that the presence of charged dusts as well as the increase of the negative ion density decrease the   amplitude of shocks. This is expected as the introduction of charged dusts causes dissipation via the collisional term and the viscosity term. Also, as the negative ion density increases, more electrons will be removed by the dust grains in order to maintain the charge neutrality. This also results to the increase of the positively charged dust density and hence may cause   the reduction (in magnitude) of the wave amplitude.

Figure 8 shows the developments of solitary waves into shocks for plasmas with negatively charged dusts.  {In absence of the dissipative effects [Fig. 8(a)], a different wave pattern with a   train of solitons is seen to form due to the dominant role of wave dispersion over the nonlinearity. As the dissipation due to viscosity effects starts playing a role, the number of wave trains    reduces and the wave  steepening begins to occur at a value of $\eta_j<1$ [See Fig. 8(b)]. Further increasing the value of $\eta_j$  ($\eta_j\sim1$) results into the fact that the train of   solitons  disappear leaving only one wave front with a lower value of $|\Phi|$ [See Fig. 8(c)]. From Fig. 8(d) it is seen that due to the effects of ion-dust collisions, the wave gets damped [Similar to the case of positively charged dusts (Fig. 7)] and the wave amplitude reduces in its magnitude.  We note that the damping seen in this case is qualitatively different  from the case of positively charged dusts (Fig. 7). From Fig. 8(d) we also find that in contrast to Fig. 7(c) where the negative ion density decreases the  wave  amplitude,  as the  number density of positive ions increases, the magnitude of $\Phi$ increases. Here the increase of the positive ion density causes the addition of more electrons into the dust grain surface, which, in turn, increases   the dust charge state in order to maintain the charge neutrality. As a  result, the dissipative effects due to dust density enhancement become more pronounced and the wave steepening occurs with an   increased $|\Phi|$.}    
\begin{figure*}
\includegraphics[height=4in,width=6in]{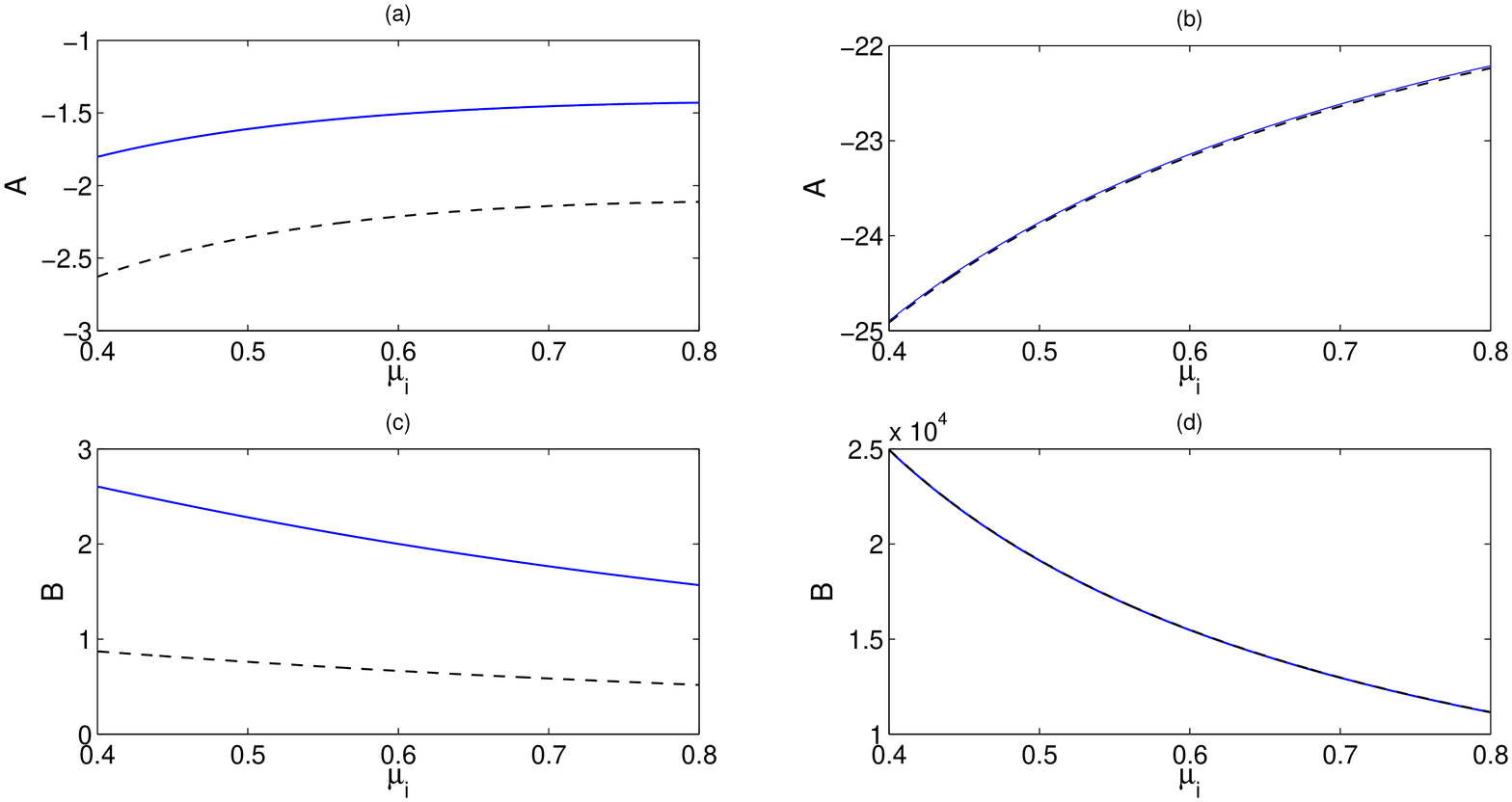}  
\caption{(Color online) The plots of the nonlinear coefficient $A$ (upper panels) and the dispersion coefficient $B$ (lower panels) versus the positive to negative ion density ratio $\mu_i$ are shown for plasmas with positively charged dusts. The other parameters are  as in Fig. 1. The solid and dashed lines correspond to $T_n=T_e/8$, $T_p\sim T_e$  and $T_n=T_e/10$, $T_p\sim T_e/2$   respectively. In the subplots [(a), (c)] and [(b), (d)], the values of  $v_0^2$  [Eq. \eqref{phase-velo}] corresponding respectively to the negative and positive signs    have been considered  to compare the results as well as to show that  the numerical values of $A$ and $B$ in the subplots [(a), (c)] may favor (while the others may not) the formation of solitary waves and shocks with negative potentials. } 
\end{figure*}
%%%%%%
\begin{figure*}
\includegraphics[height=4in,width=6in]{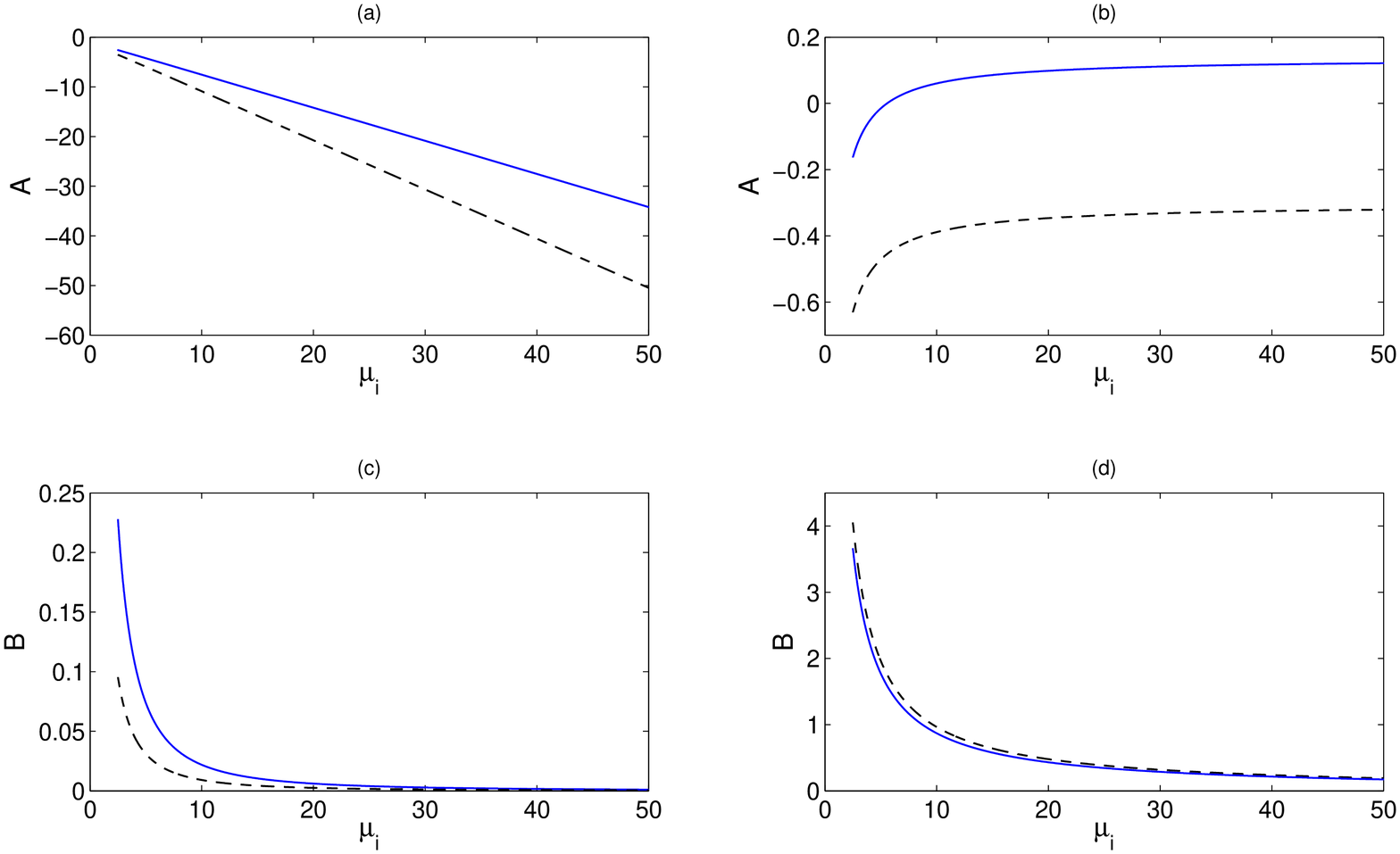}  
\caption{(Color online) The plots of the nonlinear coefficient $A$ (upper panels) and the dispersion coefficient $B$ (lower panels) versus the positive to negative ion density ratio $\mu_i$ are shown for plasmas with negatively charged dusts. The other parameters are  as in Fig. 1.   The solid and dashed lines correspond to $T_n=T_e/8$, $T_p\sim T_e$  and $T_n=T_e/10$, $T_p\sim T_e/2$   respectively. In the subplots [(a), (c)] and [(b), (d)] the values of  $v_0^2$  [Eq. \eqref{phase-velo}] corresponding respectively to the negative and positive signs    have been considered  to compare as well as to show that  the numerical values of $A$ and $B$ in the subplots [(b), (d)] may favor (while the others may not) the formation of solitary waves and shocks with both positive and negative potentials.   } 
\end{figure*}
 
 {Another interesting feature   may be the formation of double-layer solutions of Eq. \eqref{KdVB-eq}. Such solutions consist of two layers with positive and negative space charge and support localized electric fields. Also, double-layers (DLs) can play important roles in providing a mechanism for supporting electric fields in collisionless plasmas with nearly zero resistivity. Furthermore, DLs have relevance to charge particle acceleration in cosmic plasmas and in plasma thrusters. However, there are several mechanisms by which DLs can be formed (See, e.g.,  Ref. \cite{Singh-DLs} for some recent review works on DLs). Nevertheless, such DLs are more difficult to generate in a laboratory and require a fine tuning of the plasma parameters. However, the detail discussion on the formation and the properties of DLs is limited to the  present study as we have mainly focused on the formation and the characteristics features of  solitons, and shocks so generated  due to dissipative effects. } 

 {In what follows, Eq. \eqref{KdVB-eq} can have a travelling wave solution, in absence of the collisional effects $(\nu=0)$, given by \cite{21-Book-analytic-solution}
\begin{eqnarray}
&&\Phi(\xi,\tau)=C_1-\frac{12\eta^2}{25AB\left(1+C_2e^{\psi}\right)^2},\notag\\ &&\psi=-\frac{\eta}{5B}\xi+\left(\frac{\eta AC_1}{5B}-\frac{6\eta^3}{125AB^2} \right)\tau, \label{analytic-sol}
\end{eqnarray}
where $C_1$ and $C_2$ are arbitrary constants. The profiles of Eq. \eqref{analytic-sol} for different plasma parameters are shown in Fig. 9. Figures 9(a)-(c) exhibit some shock profiles  corresponding to plasma parameters relevant for negatively charged dusts, whereas Fig. 9(d) is for negative ion plasmas with positively charged dusts.   We find that the DLs with both the compressive and rarefactive potentials may exist [Figs. 9(a) and (b)] for a certain range of  values of $\eta_{j0}$, beyond which DLs with only negative  potential (rarefactive) may be formed [Fig. 9(c)]. From Figs. 9(a) and (b), it is also seen  that for a fixed negative ion density as the  density of positive ions increases, the value of
 $|\Phi|$ increases by the same physical reason as in Fig. 8(d). The opposite trend may be seen to occur by increasing the negative ion density (in   case of positively charged dusts) or by decreasing the values of $\eta_{j0}$. The effects of the latter are   shown in Fig. 9(c) exhibiting DLs with only negative potential. It turns out that  the wave front with two layers may not be formed by  gradually decreasing the values of $\eta_{j0}$ or by reducing  the charged dust state or density in negative ion plasmas with negatively charged dusts, rather we can  see the shock wave fronts with only negative potentials. Figure 9(d) also shows that if the dusts are positively charged or electrons are almost absorbed by the dust grains in pair-ion plasmas, the formation of DLs may not be possible. However, a shock profile with negative potential may exist with a significant drop of the wave amplitude $|\Phi|$ compared to the case of negatively charged dusts. }

\begin{figure*}
\includegraphics[height=4in,width=6in]{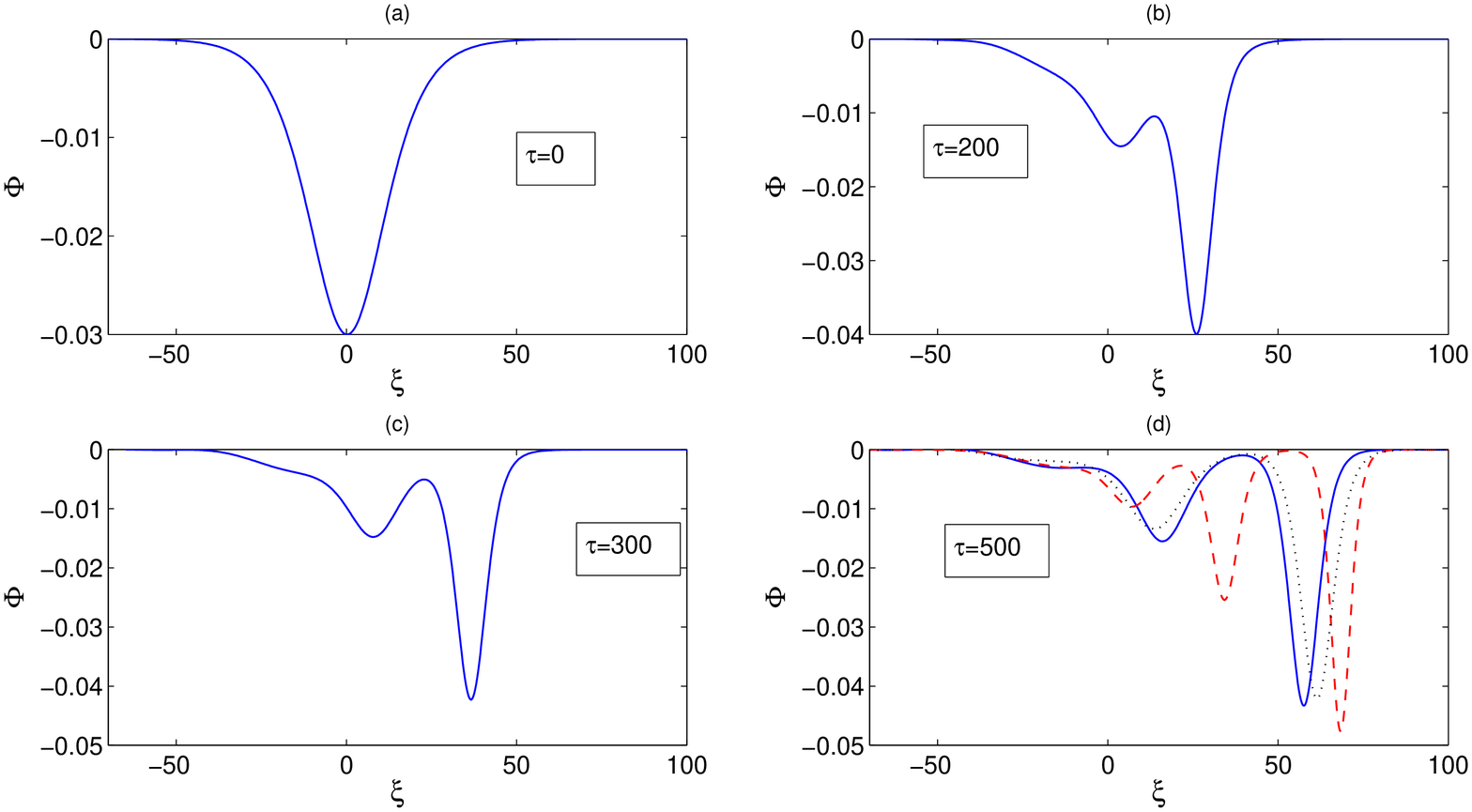}  
\caption{(Color online) Development of an initial pulse of the form $\Phi(\xi)=-0.03\text{sech}^2(\xi/15)$ into solitary waves at different times for the parameters as in Fig. 1, and when $\tilde{\nu}_{jd}=\tilde{\eta}_{jd}=0$. In the subplot (d), the dashed (red) and dotted (black) lines correspond to the cases of no dust   and   when negative ion density is different, i.e., $n_{n0}=1000n_{e0}$ in the plasma respectively. } 
\end{figure*}
%%%%%
\begin{figure*}
\includegraphics[height=4in,width=6in]{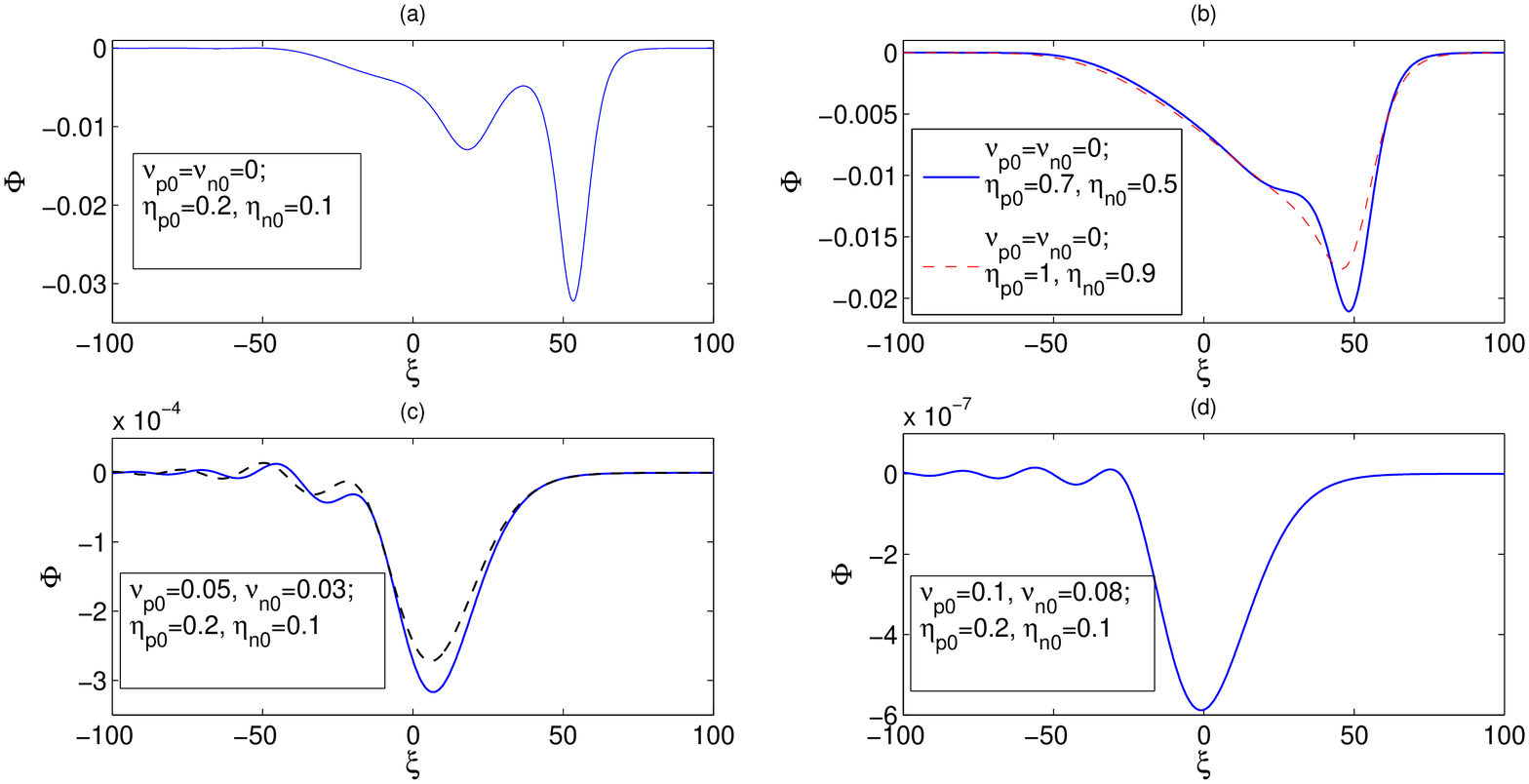}  
\caption{(Color online) Development of an initial pulse of the form $\Phi(\xi)=-0.03\text{sech}^2(\xi/15)$ into shocks   at $\tau=500$ for different values of $\nu_{j0}$ and $\eta_{j0}$ as in the figure. The   parameters values are as in Fig. 1. In the subplot (c), the dashed (black) line corresponds to the case of different negative ion density $n_{n0}=900n_{e0}$ with the same other parameters as the solid line. } 
\end{figure*}
%%%%
\begin{figure*}
\includegraphics[height=4in,width=6in]{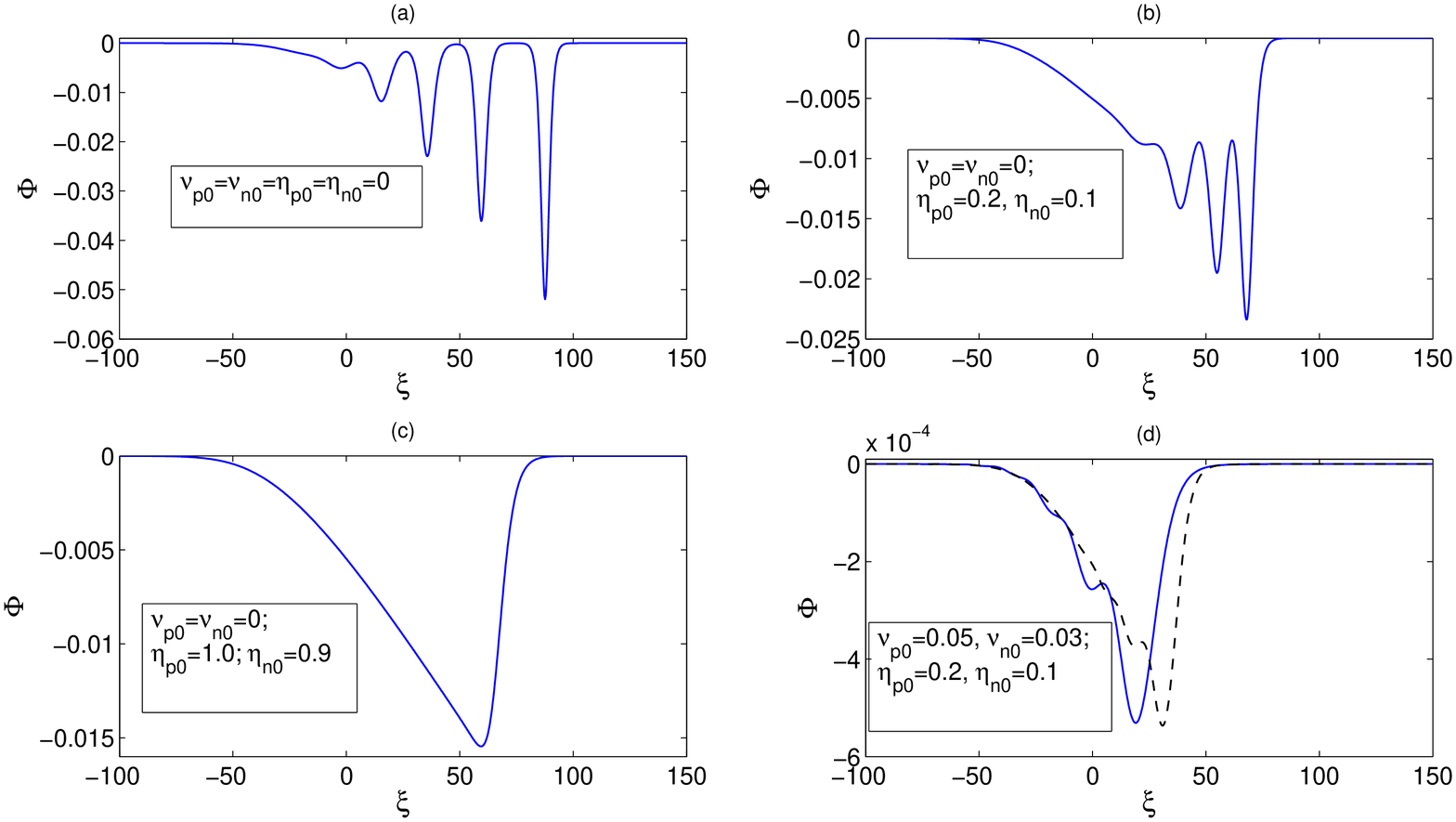}  
\caption{(Color online) Development of an initial pulse of the form $\Phi(\xi)=-0.03 \text{sech}^2(\xi/15)$ into shocks  at $\tau=500$ for different values of $\nu_{j0}$ and $\eta_{j0}$ as in the figure. The parameters are for plasmas with negatively charged dusts where $n_{e0}=n_{n0}/10$, $n_{p0}=20n_{e0}$. Others parameters are as in Fig. 1. In the subplot (d), the dashed (black) line corresponds to   different value of the positive ion density, i.e., $n_{p0}=40n_{e0}$  with the same other parameters as the solid line.  } 
\end{figure*}
%%%%
\begin{figure*}
\includegraphics[height=4in,width=6in]{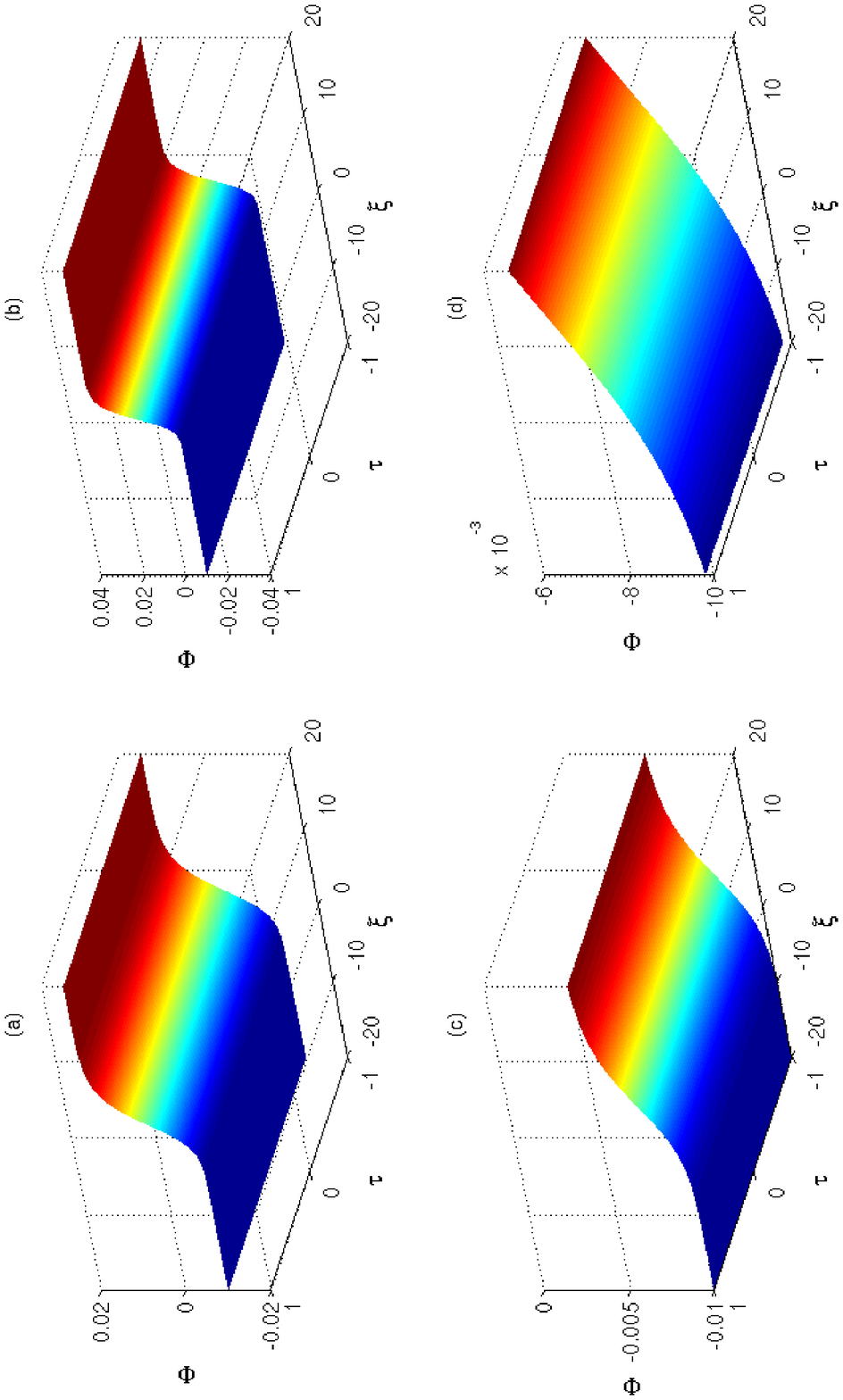}  
\caption{(Color online) Analytic solution of Eq. \eqref{analytic-sol} for different plasmas with negatively [subplots (a)-(c)] and positively charged dusts [subplot (d)]: (a) $n_{e0}=n_{n0}/2$, $n_{p0}=5n_{e0}$, $\eta_{p0}=1.2$, $\eta_{n0}=1.1$; (b) $n_{e0}=n_{n0}/2$, $n_{p0}=10n_{e0}$, $\eta_{p0}=1.3$, $\eta_{n0}=1.2$; (c) $n_{e0}=n_{n0}/2$, $n_{p0}=5n_{e0}$, $\eta_{p0}=0.7$, $\eta_{n0}=0.6$; (d) $n_{e0}=n_{n0}/700$, $n_{p0}=500n_{e0}$,    $\eta_{p0}=1.2$ and $\eta_{n0}=1.1$. } 
\end{figure*}

To summarize, we have investigated the propagation characteristics of DIA waves in a collisional negative ion plasma with immobile charged dusts. The latter may be positively charged when the number density of negative ions exceeds that of positive ions and is much larger than that of electrons. In the linear regime, we have studied numerically the ion-acoustic instability with plasma parameters relevant for both laboratory \cite{19-Kim-Merlino,PSS} and space plasmas \cite{20-Rapp}.    We find that the two modes, namely `fast' and `slow' waves exist in dusty negative ion plasmas.  Due to the higher values of the phase velocity of the fast ion wave than the ion thermal speeds, the Landau damping effect on dissipation is negligible, however, the damping of the solitary waves is mainly caused by the ion-dust collisions and the ion kinematic viscosities.  {We find that the slow modes propagate without any instability and with frequency below the negative ion plasma frequency.   However,  for long wavelength fast modes,} the damping effect is mainly due to the ion-dust collisions. For   laboratory negative ion plasmas when the electron flow is sufficient to drive the ion-acoustic instability, the inclusion of micron seized dust grains may damp the instability due to their collision with ions.  The frequency and the growth rates are reduced when dusts are positively charged. However, the opposite trend occurs when dusts are negatively charged. Thus, the phase velocity of the waves may decrease or increase with the dust density according to when the dusts are positively or negatively charged. These results suggest a  possible di1agnostic for the presence of positively or negative charged dusts as well as for the effects of ion-dust collisions in dusty negative ion plasmas \cite{19-Kim-Merlino,PSS}.   For the space situation, though the behaviors remain similar, however, the wave frequency and the instability rate increase faster with the wave number than the case of  laboratory plasmas. 

We have also studied the nonlinear propagation of small amplitude fast modes as DIA SWS in  dusty negative ion plasmas. We show that the evolution of such waves can be described by a modified  KdVB equation. The latter generalizes and modifies the previous investigation in negative ion plasmas \cite{12}. The KdVB equation is numerically solved to show that the perturbations with negative potential may propagate as SWS in plasmas with positively charged dusts, whereas SWS with both positive and negative potential may exist when dusts are negatively charged. The profiles of these SWS with only negative potential  are shown graphically (as those with positive potential are similar), and analyzed with parameters relevant for laboratory and space plasmas. We find that, in contrast to the effects of negatively charged dusts,  the presence of positively charged dusts   reduces the wave amplitude, but   enhances the width of solitary waves. 
The theoretical results may be useful for the observation of dust ion-acoustic waves in space plasmas, e.g., a dusty meteor trail region in the upper atmosphere as well as the experimental verification of the excitation of ion-acoustic instability and the nonlinear propagation of ion-acoustic solitary and shock waves in dusty multi-ion plasmas.  

There may be some open issues, e.g., the theoretical deduction of ion kinematic viscosity and the effects of ion drag forces due to positive and negative ions on the dust particles could be problem of interest but beyond the scope of the1 present investigation. Such forces, which are opposite in direction, may also be comparable in magnitude when the positive and negative ion densities are comparable. Since the   outward ion drag force is known to be responsible for the formation of a void under microgravity conditions, the presence of both the positive and negative ions could forbid the formation of voids.  

\section*{acknowledgments} This work was partially supported by  the SAP-DRS (Phase-II), UGC, New Delhi, through sanction letter No. F.510/4/DRS/2009 (SAP-I) dated 13 Oct., 2009, and by the Visva-Bharati University, Santiniketan-731 235, through Memo No. Aca-R-6.12/921/2011-2012 dated 14 Feb., 2012.

%References

%%17. A. Rahman, F. Sayed and A. A. Mamun, Phys. Plasmas 14, (2007) 034503.

\end{document}